# Magnetic nanocomposites: new methodology for micromagnetic modeling and SANS experiments

Sergey Erokhin[1], Dmitry Berkov[1], Nataliya Gorn[1], and Andreas Michels[2]

[1]*INNOVENT Technology Development*, Jena, Germany
[2]Laboratory for the Physics of Advanced Materials, University of Luxembourg, Luxembourg

## ABSTRACT

A new methodology for micromagnetic simulations of magnetic nanocomposites is presented. The methodology is especially suitable for simulations of two-phase composites consisting of magnetically hard inclusions in a soft magnetic matrix phase. The proposed technique allows to avoid the unnecessary discretization of the 'hard' inclusions (these are normally in a single-domain state), but enables an arbitrary fine discretization of the 'soft' phase. The method is applied to the determination of the equilibrium magnetization state of an iron-based nanocomposite from the Nanoperm (FeZrBCu) family of alloys and to the calculation of the corresponding small-angle neutron scattering (SANS) cross-section. For this highly interesting material, the results of our simulations exhibit a remarkable agreement with the non-trivial 'clover-leaf' SANS cross-sections observed experimentally (A. Michels et al., *Phys. Rev. B*, **74**, 134407, 2006).

**PACS:** 75.75.+a, 75.40Mg, 61.12.Ex, 75.50.Tt

# I. INTRODUCTION

Magnetic nanocomposites are bulk materials which consist of magnetic nanocrystals that are embedded in an amorphous, usually magnetically soft phase (matrix). The growing interest in this class of magnetic materials is caused by their non-trivial magnetic properties, which are highly interesting both from the fundamental point of view and for existing and potential applications of such nanocomposites. In particular, the magnetic microstructure in nanocrystalline ferromagnets can be highly inhomogeneous, which is mainly a consequence of the following two microstructural features. First, magnetic material parameters (saturation magnetization $M_S$, exchange-stiffness constant $A$, anisotropy constant $K$) for the two constituent magnetic phases may be very different. Second, at each grain or phase boundary the crystallographic anisotropy axes change their directions randomly, thus altering the locally preferred magnetization directions. In addition, new physical effects may be expected whenever the structural 'correlation length' of the microstructure − in this case the average grain size $D$ − is reduced below a characteristic intrinsic magnetic length scale which is linked to the atomistic origin of magnetism ('crossing length scales scenario'). An outstanding example is the phenomenon of exchange softening [1] that is observed in Fe-based alloys whenever $D$ becomes smaller than the so-called exchange correlation length $l_K \sim (A/K)^{1/2}$, which for Fe-based alloys is about 20-40 nm.

The inherently existing nonuniformity in the spin system leads to very interesting magnetic properties of nanocrystalline magnets, which can be sometimes very different from the corresponding features of their coarser-grained (microcrystalline) counterparts. Famous examples for alloy development (based on the above sketched rationales) include nanocrystalline NdFeB based permanent magnets [2, 3, 4, 5] or Fe-based alloys of the Finemet (Vitroperm), Hitperm or Nanoperm type [1, 6, 7], which are magnetically extremely soft and due to their high permeability are widely used as transformer cores and shielding material.

The downscaling of the individual nanosized building blocks poses increasing demands on observational techniques to resolve ever finer details of the magnetic microstructure. Most commonly used techniques such as Kerr and Lorentz microscopy [8], magnetic force microscopy [9], spin-polarized scanning tunneling microscopy [10, 11], or photoelectron spectroscopy [12] generally image the magnetic microstructure at the sample *surfaces*. In addition, in order to probe structure from macroscopic dimensions down to the atomic scale one has to combine all these techniques.

In contrast to the above mentioned methods, magnetic small-angle neutron scattering (SANS) is probably the only technique that is capable of studying the spin distribution in the *volume* of a magnetic material *and* on a length scale from several nanometers up to a few hundreds of nanometers (for recent reviews on magnetic SANS, see Refs. [13, 14, 15]).

Magnetic SANS is also a versatile technique, which allows one to investigate a wide range of materials, including ferrofluids, magnetic nanoparticles in a polymer matrix, magnetic recording media, colossal magnetoresistance materials, superconductors, spin glasses, amorphous metals, Invar alloys, magnetic single crystals, molten and solid elemental ferromagnets (Fe, Co, Ni, Tb, Gd), nanowires, precipitates in steels, diluted paramagnets in deuterated solutions, etc. [13, 14, 15, 16, 17, 18, 19]. Nanocomposite materials have been also extensively studied using SANS, whereby both magnetically hard [20, 21] and soft [18, 22, 23, 24, 25, 26, 27, 28, 29, 30] systems have been investigated and a couple of interesting new results have been obtained. For instance, studies of FeSi-based nanocrystalline soft magnets by this technique have identified layers of reduced magnetization at the interfaces between the FeSi crystals and the surrounding amorphous matrix [22, 23, 24]. Also, a transition to a superparamagnetic behavior at the ordering temperature of the matrix phase was found [26]. Magnetic-field dependent SANS measurements in combination with Kerr microscopy on magnetically textured FeSi-based ribbons [27, 28] have revealed the domain orientation and the characteristic length scale of intradomain spin misalignment.

In contrast to nuclear SANS, where the theoretical concepts are relatively well established [31], the understanding of magnetic SANS on bulk ferromagnets is still at the beginning. The main difficulty comes from a variety of competing interactions present in a ferromagnet. In order to analyze magnetic SANS data on magnetic materials in general and in magnetic nanocomposites in particular, one should be able to calculate the corresponding equilibrium magnetization configuration. The most widely used mesoscopic theoretical approach for this purpose is the micromagnetic phenomenology [32, 33, 34], where four main contributions to the total magnetic free energy (external field, magnetocrystalline anisotropy, exchange and magnetodipolar interaction energy) are taken into account. Determination of an equilibrium magnetization state of a ferromagnet using the micromagnetical framework amounts to the solution of a set of nonlinear partial differential equations, which can not be done analytically in most practically relevant problems.

Therefore, closed-form expressions for the ensuing so-called spin-misalignment scattering cross-section are limited to the approach-to-saturation regime, where the micromagnetical equations can be linearized. Pioneering work in this direction was carried out in [35], where the magnetic SANS cross-section arising from spin disorder due to dislocations in ferromagnetic metals was computed. The ansatz from [35] was generalized by Michels and Weissmüller [15], who developed the theoretical framework for analyzing random-anisotropy-type nanocrystalline ferromagnets. However, the usefulness of these analytical theories is limited due to the difficulties mentioned above.

A semianalytical model for the interpretation of SANS measurements on nanostructured simple metals presented in [36] assumes that the magnetic material consists of *spherical domains* embedded into a *homogeneously* magnetized matrix. The magnetization orientation inside each domain was determined assuming that each domain possesses a uniaxial anisotropy with the value computed from the random-anisotropy model of Herzer [1, 7] (domains were supposed to consist of several magnetically coupled grains). This (Stoner-Wohlfarth-type) model could explain qualitatively the evolution of the SANS intensity contours with the applied field in nanocrystalline Fe and Co. However, due to a very simplified treatment of the interaction between the 'domains' (neglecting, e.g., the magnetodipolar interaction) and between the 'domains' and the matrix this model can not be applied to the physically most interesting case of an inhomogeneously magnetized matrix and thus is not suitable for quantitative studies of remagnetization processes in multi-phase composites.

For this reason, full-scale numerical micromagnetic studies of magnetic SANS are clearly necessary. In the recent decade, such simulations have become possible due to the steady increase in the computer power and due to the extensive development of micromagnetic software packages such as OOMMF [37], LLG [38], MicroMagus [39] or MuMax [40]. For example, Ogrin et al. [41] used the OOMMF code to compute the magnetic SANS cross-section of CoCrPtB based longitudinal recording media. Saranu and coworkers [42] utilized the same OOMMF tool to study the effect of the magnetostatic energy and average crystallite size on the magnetic SANS of nanocrystalline ferromagnets (see Sec. II for more details).

However, all these commercially or freely available micromagnetic packages have serious drawbacks regarding the ability to simulate magnetization structures in bulk magnetic nanocomposites, as explained in detail in Sec. II below. For this reason simulations of the magnetization distribution in composite system of practical interest have been hardly possible up to now [43, 44, 45], so that a completely new methodology for micromagnetic studies of nanocomposites is required. Having such a methodology at our disposal would allow a much better understanding of equilibrium magnetization states and magnetization dynamics in nanocomposites and a deeper interpretation of corresponding results obtained via magnetic SANS. This paper is devoted to the development of such a new methodology and to the analysis of some non-trivial cross-section patterns observed recently in alloys of the Nanoperm type [18].

The paper is organized as follows: in Section II we explain why the two main classes of existing micromagnetic methods are not really suitable for simulations of hard-soft nanocomposites. In Section III we discuss our new methodology, describing the mesh-generating algorithm and the evaluation of the energy contributions, and present the results of analytical and numerical tests of our code. Section IV starts with our simulation results for equilibrium magnetization configurations in nanocomposites of the Nanoperm type. Then we compare in detail magnetic SANS cross-sections calculated numerically from these equilibrium magnetization configurations with those observed experimentally in [18]. Section V summarizes the main results obtained in this study.

## II. NON-APPLICABILITY OF EXISTING NUMERICAL MICROMAGNETIC METHODS

Micromagnetics is a mesoscopic phenomenological theory which allows - in its quasistatic version - to determine the equilibrium magnetization configuration of a ferromagnetic body, when the applied field, the geometry of the ferromagnet and its magnetic material parameters are known [32, 33, 34]. For this purpose the total magnetic free energy of a ferromagnet (which contains in the most common case contributions from the energy due to an external field, exchange, anisotropy and magnetodipolar interaction energies) is minimized. Mainly due to the non-locality of the magnetodipolar energy most practically interesting tasks can not be solved analytically, so that numerical simulations should be carried out. At present, numerical micromagnetics is a large and continuously expanding research field, which has been extensively reviewed, e.g., in the recent handbook [46]. For this reason in this subsection we briefly discuss only methodological details relevant for numerical simulations of nanocomposite materials.

First of all, we point out that such materials are probably the most complicated objects from the point of view of numerical simulations, because they consist of at least two phases with non-flat boundaries between them - a typical example is a hard-soft nanocomposite 'made of' magnetically hard grains surrounded by a 'soft' (but also ferromagnetic) matrix. Such a system is very difficult to simulate for the following reasons.

All micromagnetic simulation methods can be roughly divided into two classes - finite-difference and finite-element algorithms [46]. In the former (finite-difference) methods the system under study is discretized into a regular (usually rectangular) grid. Such a grid allows, first of all, the usage of simple finite-difference formulas for the exchange field calculation, which in the continuous formulation is the differential operator acting on the magnetization field $\mathbf{M(x)}$. Second, the translational invariance of a regular grid enables the application of the fast Fourier transformation (FFT) for the evaluation of the magnetodipolar interaction field (energy). For a system discretized into $N$ finite elements the FFT technique reduces the operation count for this nonlocal interaction from $\sim N^2$ to $\sim N \log N$. The disadvantage of a regular grid is a pure approximation for the curved boundaries. This is a serious drawback for simulations of nanocomposites, where the accurate representation of the interphase boundaries (and associated exchange and magnetodipolar effects) is crucially important for a proper system description.

The second group of numerical micromagnetic methods (finite-elements) employs the discretization of the studied body into tetrahedrons of arbitrary shapes and sizes. Such flexible discretization allows for a quite accurate representation of curved boundaries, including those between magnetically hard inclusions and a soft matrix. The tributes to pay for this convenience are (i) complicated methods for the evaluation of the exchange field (requires a representation of a differential operator on an irregular lattice) and (ii) the impossibility to use FFT for the computation of the magnetodipolar field. This fact forces the usage of highly complicated methods for the calculation of this field, based on the decomposition of the scalar or vector magnetic potentials and solution of corresponding Poisson equations for these potentials on an irregular grid [46]. This technique can be applied to systems with open boundary conditions only, thus resulting in another limitation of the finite-element method: periodic boundary conditions (which are routinely applied in simulations of bulk materials in order to eliminate strong finite-size effects) can not be used. The impossibility to use periodic boundary conditions is a serious disadvantage when simulating SANS experiments on nanocomposites (whereby the scattering intensity is sensitive to magnetization fluctuations in the *bulk*), because surface demagnetizing effects might be very significant due to (i) a substantial volume fraction occupied by a 'soft' ferromagnetic or superparamagnetic nanocomposite matrix and (ii) a relatively small simulation volume affordable even for modern computers.

Another unfavorable feature of the tetrahedral discretization is the necessity to discretize into tetrahedrons also the hard magnetic grains, even when it is clear that the magnetization configuration within a single grain is nearly collinear. This leads to a significant increase of the number of finite elements and in the corresponding increase of the computation time (see [47] for the corresponding discussion and a suggestion how this problem might be solved in frames of the finite-element method).

Due to the reasons explained above micromagnetic modeling of SANS experiments on nanocomposites is very rare. Full-scale micromagnetic simulations of SANS measurements on a two-phase system have been reported also very recently in [41], where the authors have modeled the magnetization structure of a longitudinal magnetic recording media film. Based on experimental characterization of such media, Ogrin et al. [41] have built a two-phase model of magnetic grains consisting of a magnetically hard grain core and an essentially paramagnetic shell (although with a very high susceptibility). The authors have used the OOMMF code employing the standard finite-difference scheme, so that a very fine discretization grid (0.3 x 0.3 x 0.3 nm$^3$ cells) was necessary in order to reproduce the spherical shape of grain cores with a significant accuracy. For this reason only a limited number of grains (~ 50) could be simulated. In addition, the exchange interaction between the grains and within the matrix (representing by the merging grain shells) was neglected. Under these simplifying assumptions and using several adjustable parameters the authors could achieve a satisfactory agreement of the simulated SANS intensity profile with experimental data, showing great potential of micromagnetic simulations for interpreting the SANS experiments.

The overview presented here clearly shows that further numerical studies - including the development of new simulation methods - in this direction are highly desirable.

## III. NEW MICROMAGNETIC METHODOLOGY

### A. Mesh generation

For numerical simulations of two-phase nanocomposites described above we aim to generate a polyhedron mesh with the following properties: (a) it should allow to represent each 'hard' nanocrystallite as a single finite element, because the magnetization inside such a 'hard' grain is nearly homogeneous, (b) the mesh should allow an arbitrary fine discretization of the 'soft' magnetic matrix, and (c) the shape of meshing polyhedrons should be as close as possible to the spherical one, in order to ensure a good quality of a dipolar approximation for the calculation of the magnetodipolar interaction energy.

A mesh of polyhedral finite elements satisfying these requirements can be generated using two kinds of methods. The first group of these methods employs various modifications of a purely geometrical iterative algorithm suggested in Ref. [48] for obtaining the random close packing of hard spheres. The initial distribution of sphere centers is completely random. Then at each step the worst overlap of two spheres is eliminated by moving these spheres along the line connecting their centers until these spheres are separated. This procedure may introduce new overlaps, but they are eliminated during the next steps, so that the packing quality improves (on average). The algorithm is robust and produces the random close packing of non-overlapping spheres with any desired accuracy (see [48] for details). Unfortunately, the computation time for this method is $\sim N^2$, where $N$ is the number of elements, so that the maximal number of spheres which can be positioned by this method within a reasonable computation time is $N \sim 10^4$.

To generate a mesh with a much larger number of elements ($N > 10^5$), we have developed a 'physical' algorithm, which is based on the model of spheres interacting via a short-range repulsive potential

$$U_i = \sum_{j=1}^{N} A_{\text{pot}} \exp \left\{ -\frac{d_{ij} - (r_i + r_j)}{r_{\text{dec}}} \right\}, \tag{1}$$

where $A_{\text{pot}}$ is a constant (in a typical case $A_{\text{pot}} = 10$), $d_{ij}$ is the distance between the centers of interacting spheres with radii $r_i$ and $r_j$, and $r_{\text{dec}}$ is the interaction decay radius. Again, at the beginning of the iteration procedure, we position the sphere centers randomly. Then we move these spheres according to the dissipative equation of motion resulting from the forces derived from their interaction (1). The time step for the integration of the corresponding equation of motion is adjusted so that the total system energy decreases after each step. This procedure leads also to the decrease of the overlaps between the spheres due to the repulsive nature of the potential (1). We continue to move the spheres until the maximum overlap between them does not exceed the prescribed small value (we have tested that for our purposes the minimal remaining overlap $(r_i + r_j)/d_{ij} > 0.95$ is good enough). There exist various possibilities to increase the efficiency of this algorithm. In particular, one might decrease the decay radius of the potential $r_{\text{dec}}$ (thus making the potential 'harder'), when the overlapping between spheres decreases during the sphere motion. A typical 2D example of the configuration obtained this way is shown in Fig. 1(a). We also note that due to the random spatial arrangement of spheres obtained this way we avoid possible artifacts caused by the regular placement of finite elements.

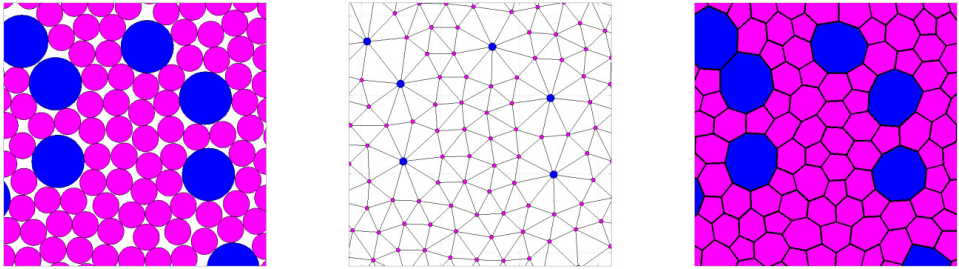

Fig. 1 (color on-line). Illustration of the algorithm for the mesh generation: (a) Random close packing of spheres having different diameters (large spheres correspond to the magnetically hard inclusions), (b) sphere centers used as location points for magnetic moments, (c) discretization of the system into the Voronoi polyhedrons corresponding to the triangulation shown in (b); these polyhedrons might be considered as finite elements used for the system discretization.

After the spheres have been positioned using one of the two algorithms described above, their centers (shown in Fig. 1(b)) are used as location points of magnetic dipoles. To compute the magnitudes $\mu_i$ of these dipoles, we first multiply the volume of the corresponding sphere by the saturation magnetization of that material inside which the dipole is located (we remind, that nanocomposites consist of materials with different magnetization). Second, we take into account that even the closely packed spheres do not fill the entire sample volume, occupying only a certain volume fraction $c_{vol}$ ($\approx 55$ % for a typical configuration) of the available space. For this reason we divide the magnitude of each dipole by this volume fraction (so that finally $\mu_i = (4\pi M_S / 3c_{vol}) r_i^3$), ensuring that in the saturated state the total magnetic moment of the discretized system is the same as by the initial continuous system.

The whole algorithm can be viewed as a method to discretize a sample into polyhedrons (see Fig. 1(c)) with a nearly spherical shape because they 'inherit' the spatial structure obtained by positioning of closely packed spheres. The fact that the shape of the volume which is 'occupied' by each magnetic moment is nearly spherical, allows us to use the dipolar approximation by the evaluation of the magnetodipolar interaction between the moments.

Finally we point out that both algorithms allow the usage of polyhedrons of different sizes, if we need different meshing on different system locations.

## B. Evaluation of energy contributions and the energy minimization

In our model we take into account the four standard contributions to the total magnetic energy: system energy in the external field, magnetic anisotropy, exchange and magnetodipolar interaction energies.

The external field and anisotropy energies (uniaxial and/or cubic) are calculated in our model in the usual way, namely

$$E_{ext} = -\sum_{i=1}^{N} \boldsymbol{\mu}_i \mathbf{H}_{ext} \tag{2}$$

$$E_{an}^{un} = -\sum_{i=1}^{N} K_i^{un} V_i (\mathbf{m}_i \mathbf{n}_i)^2 \tag{3}$$

$$E_{an}^{cub} = \sum_{i=1}^{N} K_i^{cub} V_i (m_{i,x'}^2 \cdot m_{i,y'}^2 + m_{i,y'}^2 \cdot m_{i,z'}^2 + m_{i,x'}^2 \cdot m_{i,z'}^2) \tag{4}$$

where $\mathbf{H}_{ext}$ is the external field, $\boldsymbol{\mu}_i$ and $V_i$ are the magnetic moment and the volume of the $i$-th element, and $\mathbf{m}_i$ is the unit magnetization vector. Both the anisotropy constants $K_i$ and the directions of the anisotropy axes $\mathbf{n}_i$ are site-dependent, as it is needed for a polycrystalline material. In the case of cubic anisotropy the symbols $m_{x'}$ etc. mean the components of unit magnetization vectors in the local coordinate system (attached to the cubic anisotropy axes).

For the exchange energy evaluation we use the nearest neighbors approximation. The continuous integral version of this energy employs the magnetization gradients

$$E_{exch} = \int_V A(\mathbf{r}) \left[ (\nabla m_x)^2 + (\nabla m_y)^2 + (\nabla m_z)^2 \right] dV \,, \tag{5}$$

and thus can not be discretized in a simple way, because we do not use a regular grid.

For this reason for magnetic moments belonging to the same phase we use the slightly modified form of the Heisenberg-like expression for the nearest neighbors exchange, namely

$$E_{\text{exch}} = -\frac{1}{2}\sum_{i=1}^{N}\sum_{\langle j,i\rangle}\frac{2A_{ij}\overline{V}_{ij}}{\Delta r_{ij}^2}(\mathbf{m}_i\mathbf{m}_j) \qquad (6)$$

where $\overline{V}_{ij} = (V_i + V_j)/2$, $\Delta r_{ij}$ is the distance between the dipoles $i$ and $j$ belonging to elements with volumes $V_i$ and $V_j$, and $A_{ij}$ is the exchange constant.

The sum in (6) should be performed over the nearest neighbors. The definition of these neighbors is not unique in our disordered system. We have adopted the convention that two magnetic moments are considered as nearest neighbors if they are separated by the distance not larger than $d_{\max} = 1.4\cdot(r_i + r_j)$. The factor 1.4 before the sum of the sphere radii ensures that the moments placed in the spheres which do not exactly touch each other, but which centers are located sufficiently close, are considered as nearest neighbors.

To describe the exchange between the two finite elements (polyhedrons) belonging to different phases (hard and soft), we use another formula for their exchange interaction:

$$E_{\text{ex}} = -\frac{1}{2}\sum_{i=1}^{N}\sum_{\langle j,i\rangle}\frac{2A_{ij}(V_{\text{sp}}/2)}{(\Delta r_{ij} - R_{\text{hp}})^2}(\mathbf{m}_i\mathbf{m}_j) \qquad (7)$$

Here $V_{\text{sp}}$ is the volume of a soft phase element and $R_{\text{hp}}$ is the radius of the sphere corresponding to the hard phase element. Expression (7) accounts for the fact that in this case the magnetization rotation occurs almost entirely within the polyhedron corresponding to the soft phase.

The factors in front of the scalar products $(\mathbf{m}_i\,\mathbf{m}_j)$ in (6) and (7) are proportional to the volumes of neighboring finite elements in which the magnetization rotation occurs and depend on the spacing between them. For the standard finite-difference approximation of the exchange integral (5) on a rectangular lattice these factors should be chosen so, as to assure the convergence of the sums (6) and (7) to the corresponding integral (5) for small angles between adjacent moments (see [49] for details). For a disordered system, it is necessary to correct both expressions (6) and (7) by the factor $f = 6/n_{\text{av}}$, where $n_{\text{av}}$ is the average number of neighbors in our disordered system, because 6 is the number of nearest neighbors in the 3D rectangular lattice, for which the expression (6) has been initially derived.

The last energy term - energy of the magnetodipolar interaction between moments and the corresponding contribution to the total effective field - is computed in the dipolar approximation (8),

$$E_{\text{dip}} = -\frac{1}{2}\sum_{i=1}^{N}\boldsymbol{\mu}_i\sum_{j\neq i}\frac{3\mathbf{e}_{ij}(\mathbf{e}_{ij}\boldsymbol{\mu}_j) - \boldsymbol{\mu}_j}{\Delta r_{ij}^3} \qquad (8)$$

i.e., magnetic moments of finite elements are treated as point dipoles located in the centers of generated closely packed spheres (see above). This approximation would be exact only for spherical finite elements. Hence it results in some computational errors for our discretized system, which should be considered, strictly speaking, as composed of finite elements in form of polyhedrons (as shown schematically in Fig. 1(c)). However, these errors are small, because the shape of our finite elements is close to spherical, due to the special algorithm employed for the generation of magnetic moment location points.

The summation in (8) is performed by the so called particle-mesh Ewald method. Didactically very nice and detailed general introduction into the Ewald methods can be found in [50]. The corresponding specific implementation for the magnetodipolar interaction for lattice-based and disordered systems of magnetic particles is described in our papers [51, 52]. Here we would like only to remind that the lattice Ewald method for disordered systems consists of the following stages: (*i*) mapping of magnetic moments of the initial (disordered) system onto a rectangular lattice, (*ii*) evaluation of the magnetodipolar field for the translationally invariant system obtained this way using the standard lattice Ewald method and (*iii*) backward interpolation of the field obtained on the rectangular grid on the previous step onto the initial positions of magnetic dipoles. The most time-consuming step (*ii*) can be performed using the FFT technique - due to the presence of a translationally invariant grid - which allows to reduce the operation count from $N^2$ (for the standard Ewald method) to $N·\log(N)$. Hence simulations of systems consisting of $N \sim 10^5 - 10^6$ moments become possible.

The major sources of errors in the method described above are the mapping of the initial disordered system onto a rectangular grid and the back-interpolation of the magnetodipolar field (errors for the properly implemented lattice Ewald method are vanishingly small). However, these errors may be reduced below the desired threshold by the proper choice of the mapping scheme (see [53] for the corresponding detailed description). In our case we have found that if we choose the lattice constant equal to the radius of the smallest spheres used for the mesh generation, then the usage of the linear mapping scheme together with the separate evaluation of the nearest-neighbor contribution ensures that corresponding errors are less than 1%.

To minimize the total magnetic energy, obtained as the sum of all contributions discussed above, we have used the simplified gradient method employing the dissipation term in the Landau-Lifshitz equation of motion for magnetic moments [46]. This means that the magnetization configuration at each step is updated according to the rule

$$\mathbf{m}_i^{new} = \mathbf{m}_i^{old} - \Delta t \left[ \mathbf{m}_i^{old} \times \left[ \mathbf{m}_i^{old} \times \mathbf{h}_i^{eff} \right] \right] \tag{9}$$

where $\mathbf{m}_i$ denotes the unit magnetization vector $\mathbf{m}_i = \mathbf{M}_i/M_S$ and $\mathbf{h}_i^{eff} = \mathbf{H}_i^{eff}/M_S$ is the reduced effective field $\mathbf{H}_i^{eff}$, evaluated in a standard way as the negative derivative of the total energy over the magnetic moment projections [46].

Since we are looking for the energy minimum, the time step choice in (9) is based on the monitoring of the total energy: if after the iteration step the total energy decreased, we accept this step. If the energy increased, we restore the previous magnetization state, halve the time step ($\Delta t \rightarrow \Delta t/2$) and repeat the iteration. To avoid unnecessary small time steps, we used a simple adaptive step size control: the time step is doubled, if the last few steps (typically 5 – 10 steps) were successful. The minimization process is terminated, if the maximal torque acting on magnetic moments is sufficiently small: $\max_{\{i\}} \left| \left[ \mathbf{m}_i \times \mathbf{h}_i^{eff} \right] \right| < \varepsilon$ (this condition is more sensitive than the often used criterion of the energy difference between the two subsequent steps). We have found that in all tested cases the value $\varepsilon = 10^{-4}$ was small enough to ensure the convergence of the minimization procedure.

### C. Numerical tests of the new methodology

The methodology explained above was already tested on two example problems (see the brief report in [54]). First, we have reproduced with a high accuracy the analytically known magnetization profile for a standard 3D Bloch wall. Second, for a trial 3D magnetic configuration defined via some simple trigonometric functions of coordinates (to ensure smooth spatial variations of the system magnetization) we have obtained a very good agreement between the energy values found by the new method and the micromagnetic package MicroMagus [39], which employs the standard finite-difference formalism.

In this paper we present two additional tests where we compare the equilibrium magnetization configurations of a cubic magnetic particle obtained also via our new method and the Micro-Magus package. The particle size was chosen to be $40 \times 40 \times 40$ nm$^3$, saturation magnetization was set to $M_S = 800$ G, exchange constant $A = 1 \times 10^{-6}$ erg/cm and the uniaxial anisotropy constant $K = 5 \times 10^5$ erg/cm$^3$. For the application of the new method the cube was discretized (using the algorithm described above) into $N = 9000$ elements with the typical size $d = 2$ nm. For the standard finite difference simulations the cell size $2.5 \times 2.5 \times 2.5$ nm$^2$ was used.

For the tests we have chosen two magnetization configurations which are typical for particles of this size: the vortex state and the so called 'flower' state. To obtain the vortex state, we have started the minimization procedure from the magnetization state that is topologically equivalent to the vortex – the so called closed Landau domain configuration. The flower state was obtained by starting the energy minimization from the homogenous configuration with magnetization directed along one of the cube sides.

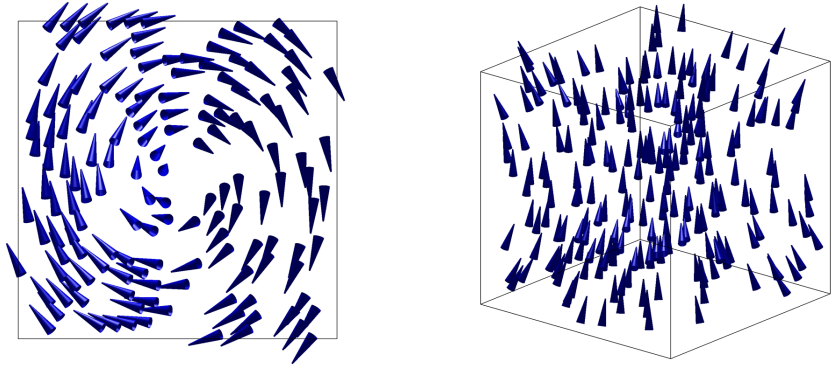

Fig. 2 (color on-line). Vortex (left image, 2D cross-section) and flower (right image, 3D plot) magnetization configurations obtained by the new method explained above in Sec. III.A and B.

|  | **Vortex** energies ($\times 10^{-11}$ erg) | | **Flower** energies ($\times 10^{-11}$ erg) | |
| --- | --- | --- | --- | --- |
|  | New method | MicroMagus | New method | MicroMagus |
| $E_{\text{tot}}$ | 8.225 | 8.270 | 7.813 | 7.843 |
| $E_{\text{an}}$ | 1.361 | 1.385 | 0.137 | 0.127 |
| $E_{\text{exch}}$ | 4.409 | 4.562 | 0.434 | 0.441 |
| $E_{\text{dem}}$ | 2.455 | 2.324 | 7.242 | 7.275 |
| $M/M_S$ | 0.400 | 0.406 | 0.972 | 0.974 |

Table 1. Comparison of energies and reduced magnetizations (last row) for the vortex and flower magnetization states by the new method and the standard finite difference simulations (MicroMagus software)

Comparison of energies for the equilibrium magnetization states (shown in Fig. 2) obtained by the new method and the standard finite difference simulations (MicroMagus package) are presented in Table 1. It can be seen that the energies obtained by both methods agree very well. The only energy exhibiting a significant relative difference is the anisotropy energy for the flower state; however, this significant relative difference ($\Delta E/E$) arises simply due to a very low value of this energy itself. All in one, the agreement between the new and established methodologies for all cases where the standard methods are applicable, are fully satisfactory.

## IV. SIMULATIONS OF MAGNETIZATION PROCESSES AND MAGNETIC SANS IMAGES FOR NANOCOMPOSITES

### A. Simulation of magnetization processes

To study the micromagnetic properties of nanocomposites, we have first performed simulations of magnetization processes for a two-phase model system. This system should imitate the Nanoperm composite studied in [18], where the hard phase consisted of Fe precipitates with the average size $D_{hard}$ = 12 nm and the volume fraction of precipitates $c_{hard} \approx 40\%$. For this reason we have chosen by the mesh generation algorithm spheres with the diameter $D_{lr}$ = 10 nm for the representation of magnetically hard grains and spheres with the much smaller diameter $D_{sm}$ = 5 nm to discretize the 'soft' matrix ($D_{lr}$ is chosen to be somewhat smaller than $D_{hard}$, because the distance between the centers of spheres generated as explained in Sec. III.A is on average somewhat larger than the sum of their radii). An example of the generated mesh is shown in Fig. 5. Further, we have used the following magnetic parameters for 'hard' and 'soft' phases: magnetizations $M_{hard}$ = 1750 G (as for bulk Fe) and $M_{soft}$ = 550 G (calculated from the average saturation magnetization of the material and parameters of Fe crystallites), anisotropy constants $K_{hard} = 4.6 \times 10^5$ erg/cm$^3$ (also a typical value for bulk Fe) and $K_{soft} = 1.0 \times 10^3$ erg/cm$^3$ (the matrix is supposed to be nearly amorphous).

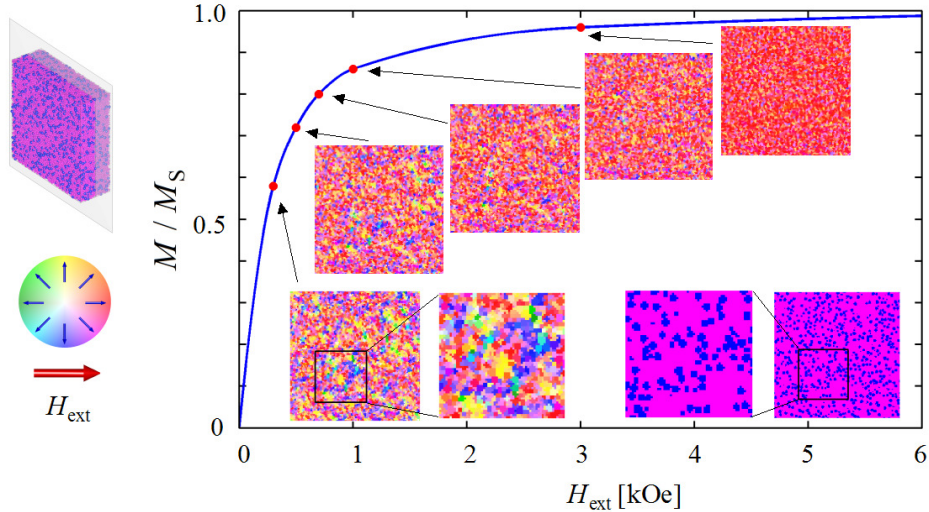

Fig. 3 (color on-line). Magnetization curve of the nanocomposite together with the color images representing the magnetization state at different fields. Magnetization states are displayed for the cross-section through our 3D simulation volume shown in the upper left corner of the figure. The inset in the lower right corner shows in blue the locations of the 'hard' inclusions which clearly correlate with the magnetization deviations from the external field direction (inset in the lower left corner).

The problem how to choose the exchange stiffness constants will be addressed in detail below. For the demo example shown in Fig. 3 we have set $A = 1.0 \times 10^{-8}$ erg/cm both for the interaction within the soft phase and between the hard and soft phases. We remind that the magnetization direction is assumed to be constant within a single Fe grain, so that there is no need to choose $A$ for the hard phase. Such a low value of the exchange constant was used to increase the saturation field value and to make the relation between the crystalline and magnetic microstructures more evident.

The size of the simulated system (rectangular box) was set to 125 x 380 x 380 nm$^3$, which was discretized into $N = 10^5$ elements. Periodic boundary conditions have been applied to avoid the effect of the stray field from the system borders.

The simulated magnetization curve together with color images visualizing the evolution of the spatial magnetization configuration is shown in Fig. 3. The color coding represents magnetic moment directions in the image plane as shown on the color wheel on the left (red corresponds to the magnetization along the applied field and green - to the opposite direction).

The main feature of the simulated magnetization process is the strong correlation between the system microstructure and the local magnetization direction. In Fig. 3 it can be clearly seen, that (in fields far from saturation) substantial magnetization deviations from the applied field direction occur on places where the 'hard' phase grains are located. This correlation is evident from the comparison of locations of 'hard' inclusions (shown in blue on the right inset at the bottom of the figure) with areas on the left inset where the magnetization substantially deviates from the field direction (visible as green-light blue-dark blue spots).

As mentioned above, a very important parameter is the exchange stiffness constant inside the soft phase and for the soft-hard phase exchange. This parameter can vary in a wide region, because it strongly depends on the concentration of iron atoms in the 'soft' phase. Unfortunately, independent measurements of the exchange constant in the soft phase are not available. A rough estimation of the exchange constant in Nanoperm ($Fe_{89}Zr_7B_3Cu$) can be done using the Curie temperatures of the similar alloy $Fe_{91}Zr_7B_2$ [55] and of bcc iron:

$$A_{Np} \sim A_{Fe} \cdot \frac{T_c^{Np}}{T_c^{Fe}} \approx 2.5 \cdot 10^{-6} \, \text{erg} / \text{cm} \cdot \frac{370 \, \text{K}}{1000 \, \text{K}} \approx 0.9 \cdot 10^{-6} \, \text{erg} / \text{cm}$$

(here we have used the literature value $A_{Fe} \approx 2.5 \times 10^{-6}$ erg/cm for the exchange constant of bcc Fe, which was computed from the data given in [56]).

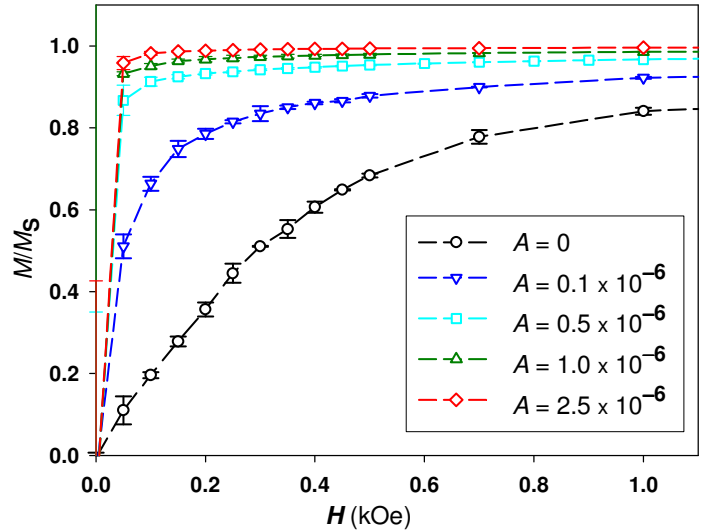

Fig. 4 (color on-line). Magnetization curves for different exchange constants $A$ for the 'soft' phase (equal to the exchange constant applied for the interaction between the 'soft' and 'hard' phases). Increase of the exchange constant leads to the rapid decrease of the saturating field value.

In addition, we have simulated the dependence of the magnetization curves on this exchange constant. In principle, corresponding results could be used to establish the value of $A$ by comparing simulated curves with those measured experimentally. Simulations have been performed for the following exchange values: $A = 2.5 \times 10^{-6}$ erg/cm, $1.0 \times 10^{-6}$ erg/cm, $0.5 \times$ $10^{-6}$ erg/cm, $0.1 \times 10^{-6}$ erg/cm for exchange interactions both within the soft phase and between the hard and soft phases and without exchange interaction at all. Simulation results are presented in Fig. 4. Each point is the result of averaging over $2 - 8$ independent realizations of the configuration of finite elements with parameters listed above. It can be seen that with increasing $A$ the system saturates faster (saturation field strongly decreases), as it should be for a system consisting of random on-site anisotropy particles interacting via a ferromagnetic exchange that is mediated by a nearly amorphous matrix. A meaningful quantitative comparison of our results with the experimental magnetization curve presented in [18] is not possible yet, because the latter contains a strong paramagnetic-type contribution (probably from the small Fe clusters in the amorphous phase) not included in our numerical model. For the comparison of our micromagnetic simulation results with the experimental neutron-scattering data (see below) we have used the value $A = 0.5 \times 10^{-6}$ erg/cm, which provides the best agreement with the experimental data.

## B. Calculation of the SANS cross-section

Most SANS experiments rely on the scattering geometry where the applied magnetic field $\mathbf{H}$ is perpendicular to the wave vector $\mathbf{k}_0$ of the incident neutron beam (compare Fig. 5). In this case, the purely magnetic elastic differential SANS cross-section $d\Sigma/d\Omega$ (for unpolarized neutrons) of a bulk ferromagnet can be expressed as [15]

$$\frac{d\Sigma}{d\Omega}(\mathbf{q}) = \frac{8\pi^3}{V} b_H^2 \left[ \left|\tilde{M}_x\right|^2 + \left|\tilde{M}_y\right|^2 \cos^2\theta + \left|\tilde{M}_z\right|^2 \sin^2\theta - (\tilde{M}_y\tilde{M}_z^* + \tilde{M}_z\tilde{M}_y^*) \cdot \sin\theta\cos\theta \right] \qquad (10)$$

Here $V$ is the scattering volume (in our simulations a rectangular prism with $125 \times 380 \times 380$ nm$^3$ and with the short side along the neutron beam, see Fig. 5), $b_H = 2.699 \times 10^{-15}$ m/$\mu_B$ ($\mu_B$ is the Bohr magneton), $a^*$ denotes a quantity complex-conjugated to $a$, $\theta$ is the angle between the external field $\mathbf{H}$ and the scattering vector $\mathbf{q} = (0,\ q \cdot \sin\theta,\ q \cdot \cos\theta)$, and $\tilde{\mathbf{M}}(\mathbf{q}) = [\ \tilde{M}_x(\mathbf{q}), \tilde{M}_y(\mathbf{q}), \tilde{M}_z(\mathbf{q})\ ]$ is the Fourier transform of the magnetization $\mathbf{M}(\mathbf{x})$. The relative contributions of the different terms (squared magnetization projections and mixed terms $\propto \tilde{M}_y\tilde{M}_z$) to the total cross-section (10) will be discussed in detail elsewhere.

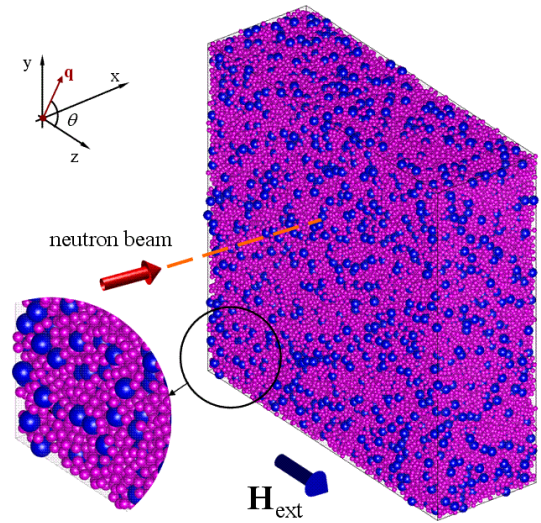

Fig. 5 (color on-line). Geometry of the neutron scattering experiment used in simulations together with the microscopic structure of the nanocomposite generated by the algorithm described in Sec. III.A (blue spheres correspond to magnetically hard grains).

In (10) we have ignored the *nuclear* SANS contribution, since the present study is devoted to simulations of *magnetic* SANS. For samples with a statistically isotropic microstructure, the nuclear coherent scattering is also isotropic and independent on the applied magnetic field. Even more important, the magnetic scattering which is relevant here due to static spin misalignment, is generally several orders of magnitude larger than the nuclear SANS signal (compare, e.g., Figs. 10, 11, 18, 22, 35 and 36 in Ref. [15]). Therefore, the simulated cross-section computed using Eq. (10) can be directly compared to experimental SANS data on nanocrystalline magnetic materials.

The Fourier components of the magnetization distribution for a disordered system can be calculated in the most efficient way by mapping (interpolating) this distribution onto a regular grid. We have used the following method: if the center of the regular grid element ($j,k,l$) is inside the $i$-th finite element of our disordered system and this element represents the 'hard' phase fraction, the corresponding magnetization vector is calculated as

$$\mathbf{M}^{hard}_{jkl} = \frac{\mathbf{m}_i M_{S,i} V_i}{N^{hard}_i},$$

where $N^{hard}_i$ is the number of regular grid elements which centers are within the $i$-th hard phase mesh element of the initial disordered system. If the center of the regular grid element is inside the soft phase, its magnetization vector is

$$\mathbf{M}^{jkl}_{soft} = \frac{\mathbf{m}_i M_{S,i} V_i}{\langle N^{soft} \rangle},$$

where $\langle N^{soft} \rangle$ is the *average* number of regular grid elements inside the finite elements used for the discretization of the 'soft' phase. Use of the average number $\langle N^{soft} \rangle$ instead of the number $N^{soft}_i$ of regular elements belonging to the $i$-th finite element helps to suppress non-physical fluctuations of the magnetization distribution, arising due to significant fluctuations of $N^{soft}_i$ for different finite elements (for computing the Fourier transform of the magnetization, we use the regular mesh which is only about two times finer than the size of a soft phase element of the initial disordered mesh).

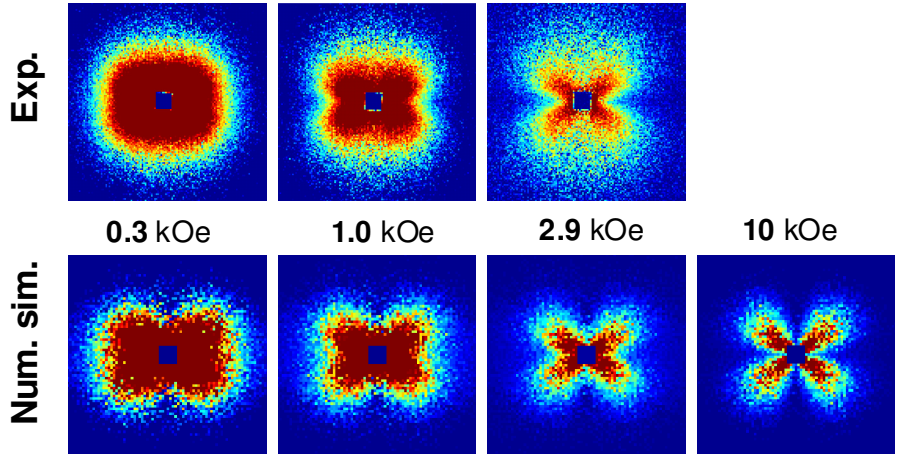

Fig. 6 (color on-line). Comparison of the SANS intensities obtained experimentally (upper row) and numerically (lower row) for different external fields as indicated (logarithmic scale for the intensities was used). Pixels in the corners of the images have momentum transfer $q = 0.64$ nm$^{-1}$. The external field is applied horizontally in the plane. Experimental data has been taken from Ref. [18].

Numerical results for the SANS cross-section obtained by evaluating the expression (10) using the algorithm outlined above are presented in Fig. 6 (lower row of images) together with the experimental results reported in [18]. Both numerical and experimental pictures represent so called difference-intensity data [18], i.e., they are obtained by subtracting the SANS cross-section in the saturated magnetization state from the cross-section obtained at each particular field. Numerical images have been averaged over 8 random realizations of the nanocomposite microstructure.

Clearly, Fig. 6 demonstrates a very good semiquantitative agreement between experimental and numerical results, including their dependencies on the external magnetic field, which is applied in the horizontal direction with respect to the images. At zero field the total (i.e., without subtracting the cross section at saturation) scattering intensity is isotropic (not shown here, see [54] for details). The scattering intensity at saturation – which is used to calculate the difference-intensity images shown in Fig. 6 – exhibits the angular dependence $\sim \sin^2\theta$, in agreement with Eq. (10): at saturation, $M_x = M_y = 0$ and $M_z = M_S$, so the only remaining term in Eq. (10) is $\sim \sin^2\theta$.

At intermediate fields (see image for $H = 2.9$ kOe), however, the SANS intensity shows a non-trivial angular anisotropy in form of the 'clover-leaf' shape, first observed experimentally in [29]. The qualitative explanation of this effect is based on the magnetization jump at the boundary between the 'hard' and soft phases, as discussed in [18]. According to this explanation, the magnetodipolar field arising due to this jump around the 'hard' inclusions, leads to the deviation of the magnetization direction in the amorphous phase from the external field direction. The angular dependence of this deviation is similar to the corresponding dependence of the stray field created by the 'hard' inclusion ($\sim \sin\theta\cos\theta$) and introduces an additional angular dependence into the SANS intensity via $\tilde{M}_x(\mathbf{q})$, $\tilde{M}_y(\mathbf{q})$, $\tilde{M}_z(\mathbf{q})$. This additional dependence is superimposed onto the trigonometric functions of $\theta$ present in Eq. (10), leading to substantial deviations of the intensity maxima locations from their trivial values $\theta = \pm 45^{\circ}$ (for $H = 2.9$ kOe the SANS image shown in Fig. 6 exhibits intensity maxima at $\theta \approx \pm 35^{\circ}$).

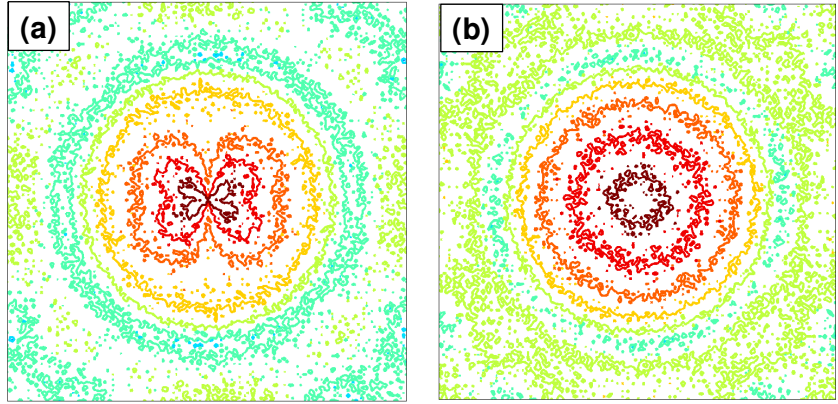

Fig. 7 (color on-line). Comparison of the angular dependencies of the Fourier component $\left|\tilde{M}_y(\mathbf{q})\right|^2$ for systems where the magnetodipolar interaction is taken into account (a) and where it is neglected (b) in the applied field $H = 2.9$ kOe. For the system without the magnetodipolar interaction (b) the clover-leaf-type pattern seen in (a) is completely absent and the angular dependence of $\left|\tilde{M}_y(\mathbf{q})\right|^2$ is fully isotropic.

To verify this explanation, we have performed simulations for the same system as discussed above, but neglecting the magnetodipolar interaction (i.e., the corresponding energy and effective field contributions have been 'switched off'). The main result of this simulation is shown in Fig. 7, where we compare the angular dependencies of the $\left|\tilde{M}_y(\mathbf{q})\right|^2$-component (computed for both systems at the same magnetization value in the applied field direction) for the system with (Fig. 7(a)) and without (Fig. 7(b)) the magnetodipolar interaction. For the system where the magnetodipolar interaction is present, the typical clover-leaf-type pattern can be clearly seen in $\left|\tilde{M}_y(\mathbf{q})\right|^2$; for certain $q$ and $H$ it delivers the main contribution to the corresponding pattern present in the total SANS cross-section. When the magnetodipolar interaction is neglected, this pattern disappears completely, thus verifying that this interaction is the main reason for the presence of the clover-leaf structure in the SANS image of magnetic nanocomposites. A more detailed discussion of the magnetic SANS of nanocomposites will be the subject of forthcoming publications.

## V. SUMMARY AND CONCLUSIONS

In this publication we have introduced a new micromagnetic methodology which is especially suitable for numerical simulations of widely used many-phase magnetic nanocomposites consisting of magnetically soft and hard phases, where the 'hard' phase inclusions have an approximately spherical shape. By applying this new approach to the simulations of magnetization distribution and subsequent calculations of the small-angle neutron scattering (SANS) cross-sections of nanocomposites of the Nanoperm type, we could achieve a very good agreement with experimental data obtained on these alloys. Our preliminary studies of the role of the various magnetic interactions in this highly non-trivial material have confirmed that the qualitatively new 'clover leaf' shape seen in experimental SANS data is indeed related to the magnetodipolar interaction and arises due to the abrupt jump in the magnetization magnitude at the boundaries between hard and soft magnetic phases, as previously suggested in [18].

**ACKNOWLEDGMENT.** This work was supported by the Deutsche Forschungsgemein-schaft under the project BE 2464/10-1 and by the National Research Fund of Luxembourg in the framework of ATTRACT Project No. FNR/A09/01 and Project No. FNR/10/AM2c/39.

## References

[1] G. Herzer, in *Handbook of Magnetic Materials*, Ed.: K.H.J. Buschow, Elsevier, Amsterdam, 1997, vol. 10, 415.

[2] J.J. Croat, J.F. Herbst, R.W. Lee and F.E. Pinkerton, *J. Appl. Phys.*, **55**, 2078 (1984).

[3] R. Coehoorn, D.B. de Mooij, J.P.W.B. Duchateau, and K.H.J. Buschow, *J. Phys.*, **49**, C8-669 (1988).

[4] E. F. Kneller and R. Hawig, *IEEE Trans. Magn.* **27**, 3588 (1991).

[5] *Magnetic Anisotropy and Coercivity in Rare-Earth Transition Metal Alloys*, Eds.: L. Schultz and K.-H. Müller, Werkstoff-Informationsgesellschaft, Frankfurt, 1998.

[6] M.E. McHenry, M.A. Willard and D.E. Laughlin, *Prog. Mater. Sci.*, **44**, 291 (1999).

[7] K. Suzuki and G. Herzer, in *Advanced Magnetic Nanostructures*, Eds.: D. Sellmyer and R. Skomski, Springer, New York, 2006, pp. 365-401.

[8] A. Hubert and R. Schäfer, *Magnetic Domains*, Springer-Verlag, Berlin, 1998.

[9] U. Hartmann, *Ann. Rev. Mater. Sci.* **29**, 53 (1999).

[10] R. Wiesendanger, *Scanning Probe Microscopy and Spectroscopy*, Cambridge University Press, Cambridge, 1994.

[11] M. Bode, *Rep. Prog. Phys.*, **66**, 523 (2003).

[12] E. Bauer, *J. Phys.: Cond. Matt.* **13**, 11391 (2001).

[13] A. Wiedenmann, *Physica B*, **356**, 246 (2005).

[14] W. Wagner and J. Kohlbrecher, in: *Modern Techniques for Characterizing Magnetic Materials*, Ed.: Y. Zhu, Kluwer Academic Publishers, Boston, 2005, pp. 65-103.

[15] A. Michels and J. Weissmüller, *Rep. Prog. Phys.*, **71**, 066501 (2008)

[16] W. Schmatz, T. Springer, J. Schelten J and K. Ibel, *J. Appl. Crystallogr.*, **7**, 96 (1974).

[17] V. Gerold and G. Kostorz, *J. Appl. Crystallogr.*, **11**, 376 (1978).

[18] A. Michels, C. Vecchini, O. Moze, K. Suzuki, P.K. Pranzas, J. Kohlbrecher and J. Weissmüller, *Phys. Rev. B*, **74**, 134407 (2006).

[19] A. Michels, *Phys. Rev. B*, **82**, 024433 (2010).

[20] A. Bracchi, K. Samwer, S. Schneider and J.F. Löffler, *Appl. Phys. Lett.*, **82**, 721 (2003).

[21] E. García-Matres, A. Wiedenmann, G. Kumar, J. Eckert, H. Hermann and L. Schultz, *Physica B*, **350**, 315E (2004).

[22] A. Danzig, A. Wiedenmann and N. Mattern, *J. Phys.: Cond. Matt.*, **10**, 5267 (1998).

[23] H. Hermann, A. Heinemann, N. Mattern and A. Wiedenmann, *Europhys. Lett.*, **51**, 127 (2000).

[24] A. Heinemann, H. Hermann, A. Wiedenmann, N. Mattern and K. Wetzig, *J. Appl. Crystallogr.*, **33**, 1386 (2000).

[25] A. Wiedenmann, *Physica B*, **297**, 226 (2001).

[26] J. Kohlbrecher, A. Wiedenmann and H. Wollenberger, *Z. Phys. B*, **104**, 1 (1997).

[27] A. Michels, R.N. Viswanath and J. Weissmüller, *Europhys. Lett.*, **64**, 43 (2003).

[28] A. Grob, S. Saranu, U. Herr, A. Michels, R.N. Viswanath and J. Weissmüller, *phys. stat. sol. (a)*, **201**, 3354 (2004).

[29] A. Michels, C. Vecchini, O. Moze, K. Suzuki, J.M. Cadogan, P.K. Pranzas and J. Weissmüller, *Europhys. Lett.*, **72**, 249 (2005).

[30] C. Vecchini, O. Moze, K. Suzuki, P. K. Pranzas, J. Weissmüller and A. Michels, *Appl. Phys. Lett.*, **87**, 202509 (2005)

[31] A. Guinier and G. Fournet, *Small-Angle Scattering of X-Ray*s, New York, Wiley, 1955; O. Glatter and O. Kratky (Eds.), *Small-Angle X-Ray Scattering*, London, Academic Press, 1982; L. A. Feigin and D. I. Svergun, *Structure Analysis by Small-Angle X-Ray and Neutron Scattering*, New York, Plenum Press, 1987; S.-H. Chen and T.L. Lin, in: *Methods of Experimental Physics-Neutron Scattering*, vol. 23B, Eds. D.L. Price and K Sköld, San Diego: Academic Press, 1987, pp. 489-543; *Neutron, X-Ray and Light Scattering: Introduction to an Investigative Tool for Colloidal and Polymeric Systems*, Eds.: P. Lindner and Th. Zemb,  Amsterdam, North Holland, 1991; J.S. Pedersen, in *Neutrons, X-Rays and Light: Scattering Methods Applied to Soft Condensed Matter*, Eds.: P. Lindner and Th. Zemb, Amsterdam, Elsevier, 2002, pp. 391-420; D. I. Svergun and M. H. J. Koch, *Rep. Prog. Phys.*, **66**, 1735 (2003); H.B. Stuhrmann, *Rep. Prog. Phys.*, **67**, 1073 (2004); Y.B. Melnichenko and G.D. Wignall, *J. Appl. Phys.*, **102**, 021101 (2007).

[32] W.F. Brown Jr., *Micromagnetics*, Interscience Publishers, New York, 1963.

[33] A. Aharoni, *Introduction to the Theory of Ferromagnetism* (Clarendon Press, Oxford, 1996).

[34] H. Kronmüller and M. Fähnle, *Micromagnetism and the Microstructure of Ferromagnetic Solids*, Cambridge University Press, Cambridge, 2003.

[35] H. Kronmüller, A. Seeger and M. Wilkens, *Z. Phys.*, **171**, 291 (1963).

[36] J.F. Löffler, H.B. Braun, W. Wagner, G. Kostorz and A. Wiedenmann, *Phys. Rev. B*, **71**, 134410 (2005).

[37] M. Donahue, D. Porter, The Object Oriented MicroMagnetic Framework (OOMMF), http://math.nist.gov/oommf/

[38] M. Scheinfein, LLG Micromagnetics Simulator, http://llgmicro.home.mindspring.com/

[39] D.V. Berkov, N.L. Gorn, Micromagus – package for micromagnetic simulations, http://www.micromagus.de

[40] A. Vansteenkiste and B. Van de Wiele, *J. Magn. Magn. Mater*, **323**, 2585 (2011).

[41] F.Y. Ogrin, S.L. Lee, M. Wismayer, T. Thomson, C.D. Dewhurst, R. Cubitt and S.M. Weekes, *J. Appl. Phys.*, **99**, 08G912, 2006.

[42] S. Saranu, A. Grob, J. Weissmüller, and U. Herr, *phys. stat. sol. (a)*, **205**, 1774 (2008)

[43] H. Fukunaga, N. Kitajima and Y. Kanai, *Materials Transactions* **37**, 864 (1996).

[44] R. Fischer and H. Kronmüller, *Phys. Rev. B* **54**, 7284 (1996).

[45] R. Fischer, T. Leineweber and H. Kronmüller, *Phys. Rev. B* **57**, 10723 (1998).

[46] *Handbook of Magnetism and Advanced Magnetic Materials*, H. Kronmüller, S. Parkin (Eds), Vol. 2: Micromagnetism, 2007, John Wiley & Sons, Ltd, Chichester, UK, pp. 795-823

[47] J. Fidler, P. Speckmayer, T. Schrefl and D. Suess, *J. Appl. Phys.*, **97**, 10E508 (2005).

[48] W.S. Jodrey and E.M. Tory, *Phys. Rev. A*, **32**, 2347 (1985).

[49] D.V. Berkov, N.L. Gorn, *Numerical simulation of quasistatic and dynamic remagnetization processes with special applications to thin films and nano-particles,* Chap. 12 of vol. II: *Simulations and Characterisation* in the '*Handbook of Advanced Magnetic Materials*', ed. by D. Sellmyer et al., Springer Verlag and Tsinghua University Press, 2005 (ISBN 7-302-08701-6/T).

[50] P. Gibbon, G. Sutmann, *Long-range interactions in many-particle simulations*, in: *Quantum simulations of many-body systems*, J. Grotendorst et al. (Eds.), Lecture Notes, Jülich, NIC Series, vol. 10, 2002

[51] D.V. Berkov, N.L. Gorn, *Phys. Rev. B*, **57** (1998)14332

[52] N.L. Gorn, D.V. Berkov, P. Görnert, D. Stock, *J. Magn. Magn. Mat.*, **310**, 2829 (2007)

[53] M. Deserno, C. Holm, *J. Chem. Phys.*, **109**, 7678 (1998)

[54] S. Erokhin, D. Berkov, N. Gorn, A. Michels, accepted in *IEEE Trans. Magn*.

[55] K.Suzuki and J.M.Cadogan, *Phys. Rev. B*, **58**, 2730 (1998).

[56] *Landolt-Börnstein: Numerical Data and Functional Relationships in Science and Technology*, vol. 19a (*Magnetic Properties of Metals*), Eds.: K.-H. Hellwege, O. Madelung, Springer-Verlag Berlin-Heidelberg, 1986, Sec. 1.1.2.9, p.74